\documentclass[a4paper]{article}
\usepackage[pdftex]{graphicx}
\usepackage{multirow,amsmath,bm,url}

\usepackage{INTERSPEECH2020}

\newcommand{\bX}{\mathbf{X}}

\newcommand{\bY}{\mathbf{Y}}

\newcommand{\bZ}{\mathbf{Z}}
\newcommand{\bz}{\mathbf{z}}

\title{Hierarchical Multi-Grained Generative Model for Expressive Speech Synthesis}
\name{Yukiya Hono$^{1,2}$, Kazuna Tsuboi$^2$, Kei Sawada$^2$, Kei Hashimoto$^1$, Keiichiro Oura$^1$, \\Yoshihiko Nankaku$^1$, and Keiichi Tokuda$^1$}
\address{
  $^1$Department of Computer Science, Nagoya Institute of Technology, Nagoya, Japan\\
  $^2$Microsoft Development Co., Ltd., Tokyo, Japan}
\email{hono@sp.nitech.ac.jp, \{ktsuboi, kesawada\}@microsoft.com, \\\{bonanza, uratec, nankaku, tokuda\}@sp.nitech.ac.jp}

\begin{document}

\maketitle

\begin{abstract}
\vspace{-1mm}
This paper proposes a hierarchical generative model with a multi-grained latent variable to synthesize expressive speech.
In recent years, fine-grained latent variables are introduced into the text-to-speech synthesis that enable the fine control of the prosody and speaking styles of synthesized speech.
However, the naturalness of speech degrades when these latent variables are obtained by sampling from the standard Gaussian prior.
To solve this problem, we propose a novel framework for modeling the fine-grained latent variables, considering the dependence on an input text, a hierarchical linguistic structure, and a temporal structure of latent variables.
This framework consists of a multi-grained variational autoencoder, a conditional prior, and a multi-level auto-regressive latent converter to obtain the different time-resolution latent variables and sample the finer-level latent variables from the coarser-level ones by taking into account the input text.
Experimental results indicate an appropriate method of sampling fine-grained latent variables without the reference signal at the synthesis stage.
Our proposed framework also provides the controllability of speaking style in an entire utterance.
\end{abstract}

\section{Introduction}
\vspace{-1mm}
Deep neural network (DNN)-based approaches have become mainstream in statistical parametric text-to-speech (TTS) synthesis in recent years~\cite{zen-2013-statistical, fan-2014-tts, wu-2016-inverstigating}.
A DNN-based acoustic model represents the mapping function between the linguistic feature sequences and the acoustic feature sequences.
Recently, end-to-end neural network-based approaches have also made significant progress~\cite{wang-2017-tacotron,shen-2018-natural,ping-2017-deep,li-2019-neural}, and the quality of synthesized speech has been greatly improved.
Most data-driven TTS synthesis approaches aim to achieve adequate neutral prosody, so synthesized speech is less expressive.
Although such an averaged voice is acceptable in short assistant-like utterance, the listener feels unpleasant when listening to such speeches in conversation.
Opportunities for humans to interact with a computer is increasing due to the spread of practical applications such as intelligent conversational agents and assistants; therefore, interest in expressive and controllable speech synthesis is increasing in the speech-synthesis research field.

A simple method of controlling speaking styles of synthesized speech is to use an additional vector such as emotion ID as the input of the acoustic model~\cite{inoue-2017-investigation}.
A conditional vector often comes from the corpus, so the variation of synthesized speech depends on the annotation.
Since making speaking-style annotation is difficult and often subjective, and the number of classes is limited, the variation and the quality of the synthesized speech are inadequate.

Recent approaches learn a latent feature from a reference speech in an unsupervised manner~\cite{skerry-2018-towards,wang-2018-style}.
With these approaches, an additional network, which is referred to as the reference encoder, is used to extract a single latent feature for an entire utterance to capture the global speech attributes of each utterance.
The latent feature can represent many speech attributes, such as speaking style, prosody, channel characteristics, and noise levels.
Variational inference~\cite{kingma-2014-auto} has also been incorporated into the TTS synthesis system~\cite{akuzawa-2018-expressive,henter-2018-deep,zhang-2019-learning}.
This offers various advantages, for example, the ability to obtain the latent variable by direct sampling from an accompanying prior without using a reference signal, and smoothly interpolating in the latent space.
However, this utterance-level variational autoencoder (VAE), where a single latent variable is extracted for each utterance, has a limitation in capturing the prosody or the speaking style at a specific moment.
A fine-grained latent variable, which is a variable-length sequence such as a word- or phone-level sequence, has been introduced into the TTS, to support sequential control of prosody and speaking style~\cite{lee-2019-robust}.
In a previous study~\cite{sun-2020-fully}, a multi-level fine-grained VAE and a dimension-wise autoregressive~(AR) decomposition of the posterior were introduced.
The authors used a conditional VAE structure to replace the original VAE.
This model conditions on the projection of latent variables, and extracts latent dimensions one at a time.

The fine-grained VAE naturally enables precise control of the speaking style of synthesized speech.
However, generated speech using latent variables by sampling from a standard VAE prior is unnatural and discontinuous, for example, the speaking style changes dramatically between units such as words and phones.
This is because the fine-grained latent variables are sampled from a Gaussian prior distribution independently for each unit such as word and phone, although an approximate posterior is derived from a reference speech and has a temporal dependence.
A recent study~\cite{sun-2020-generating} used an AR prior network conditioned on the phone embeddings, which is trained to fit the VAE posterior distribution.
Fine-grained latent variables have hierarchical linguistic and temporal dependence since the speech signal is the time-series data with a hierarchical linguistic structure.
Furthermore, the latent variables and text content should have a strong correlation.
Thus, fine-grained latent variables should be modeled and sampled by taking such dependence into account.

We propose a hierarchical multi-grained framework for modeling the fine-grained latent variables for the expressive TTS synthesis system.
Our framework consists of a multi-grained VAE, a conditional prior, and a multi-level AR latent converter.
The three different-resolution VAEs extract the utterance-, phrase-, and word-level latent variables, and the two latent converters represent the relation between the hierarchy of these variables.
We also introduce a residual structure for the encoders and the latent converters, and a decoder parameter sharing, which help in learning the hierarchical structure of latent variables.
In the synthesis phase, the word-level latent variables are predicted with the latent converters from a coarser-level latent variable that is sampled from the conditional prior.
Therefore, we can obtain the proper word-level latent variables without reference speech.
Our proposed system also provides controllability.
Since the word-level latent variables are predicted with the latent converters, we can specify the speaking style as easily as with the utterance-level VAE-based system, and generated speech is more expressive than with that system.
This is an advantage of our framework when used with certain applications.

\section{Variational autoencoder-based text-to-speech synthesis}
\vspace{-1mm}

\subsection{Variational autoencoder}
\vspace{-1mm}
A VAE~\cite{kingma-2014-auto} is the kind of deep generative model for learning complicated data distribution in an unsupervised manner.
We use a conditional VAE, which extracts the latent random variables $\bz$ to capture the variations in the observed dataset $\bX$ conditioned on the auxiliary features $\bY$.
The VAE is optimized with the evidence lower bound (ELBO) as follows:
\begin{align}
  \mathcal{L}(p, q) &= \mathbb{E}_{q(\bz \mid \bX, \bY)} [\log p(\bX \mid \bY, \bz)] \nonumber \\
  & \qquad \qquad - D_\mathrm{KL} \left( q(\bz \mid \bX, \bY) \parallel  p(\bz) \right),  \label{eq:utt-vae}
\end{align}
where the first term is the reconstruction loss and the second term is the Kullback-Leibler divergence between the prior and the posterior.
Generally, the prior $p(\bz)$ is chosen to be a centered isotropic multivariate Gaussian $\mathcal{N} (\mathbf{0}, \mathbf{I})$, and the approximate posterior $q(\bz \mid \bX, \bY)$ is a Gaussian $\mathcal{N}(\bm{\mu}, \bm{\sigma})$ with mean $\bm{\mu}$ and variance $\bm{\sigma}$ as the outputs of the neural network called an encoder.
At the training, $\bz$ is sampled from the approximate posterior.
On the other hand, during inference, $\bz$ is sampled from the prior $p(\bz)$.

\subsection{TTS synthesis system incorporating the VAE}
\label{sec:vae-tts}
\vspace{-1mm}

A VAE is applied to a TTS synthesis system by regarding $\bX$ and $\bY$ as the sequence of acoustic features and the linguistic features, respectively~\cite{akuzawa-2018-expressive,henter-2018-deep,zhang-2019-learning}.
It is noted that our experiments in this paper are based on the basic TTS framework represented by~\cite{zen-2013-statistical,fan-2014-tts} to focus on how to extract and model the latent representation for expressive speech synthesis.
Thus, the acoustic feature sequence and the linguistic feature sequence are time aligned in advance.
The encoder maps a variable-length acoustic feature to two utterance-level fixed vectors, corresponding to the posterior mean and log variance.
The decoder works as an acoustic model, which predicts acoustic features from linguistic features and latent variables.

We build the encoder and decoder with long short-term memory (LSTM)~\cite{hochreiter-1997-long}.
In the encoder, the acoustic features and the linguistic features are first passed through the fully-connected (FC) layers to ensure their dimensions equal each other.
These features are added and then fed into a stack of two bidirectional LSTM layers.
A mean pooling layer is used to summarize the LSTM outputs across time, followed by two separate FC layers with linear activation to predict the mean and log variance of the posterior distribution.
The decoder has similar architecture as the encoder, which consists of two FC layers to merge both linguistic features and latent variables, a stack of two bidirectional LSTM layers, and one FC layers to output the acoustic features.

The fine-grained VAE is a modified version of the utterance-level VAE mentioned above, to obtain variable-length latent variables such as a phone-, word-, and phrases-level latent variables.
The difference between the fine-grained VAE and the utterance-level VAE is the interval of times for summarizing the LSTM output of the encoder at the mean pooling layer.
The fine-grained VAE can be trained in the same fashion as the utterance-level VAE.

\section{Hierarchical multi-grained generative model for expressive text-to-speech synthesis}
\vspace{-1mm}
The fine-grained VAE can capture the speaking style and prosody at a specific moment, but synthesized speech using sampled latent variables from the prior directly may be unnatural.
In this section, we reconsider fine-grained latent variables and propose a novel framework that can finely control the speaking style and synthesize more expressive speech.

\subsection{Modeling speaking style}
\vspace{-1mm}

\begin{figure}[t]
  \centering
  \includegraphics[width=0.95\hsize]{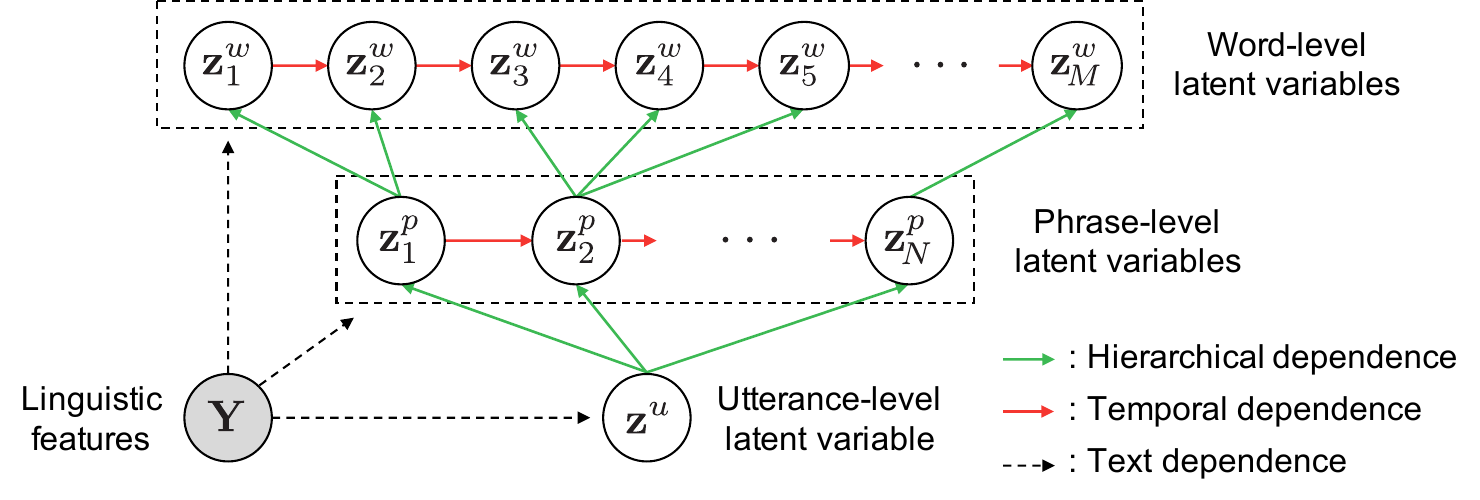}
  \vspace{-2mm}
  \caption{Graphical model of the latent variables}
  \label{fig:z_graphical}
  \vspace{-5mm}
\end{figure}

The speaking style can be factorized into multiple temporal resolution representations, such as utterance-, phrase-, and word-level representations.
These representations have a hierarchical linguistic dependency and correlate with the content of the text.
These fine-grained representations also have temporal coherency.
To incorporate these dependencies in explicitly modeling the fine-grained latent variables, we assume the graphical model of the latent variables as shown in Fig.~\ref{fig:z_graphical}.
The notation $\bz^u$ is the utterance-level latent variable, and $\bZ^p = (\bz^p_1, \bz^p_2, \cdots, \bz^p_N), \bZ^w = (\bz^w_1, \bz^w_2, \cdots, \bz^w_M)$ are the sequences of the phrase-level latent variables and the word-level latent variables, respectively, and $N$ and $M$ are the number of phrases and words in an utterance, respectively.
As shown in Fig.~\ref{fig:z_graphical}, each latent variable depends on the content of the text, hierarchical linguistic structure, and temporal structure.
The fine-grained latent variables should be sampled considering these kinds of dependencies.

\subsection{Proposed framework}
\label{sec:proposed}
\vspace{-1mm}

\begin{figure}[t]
  \centering
  \includegraphics[width=0.97\hsize]{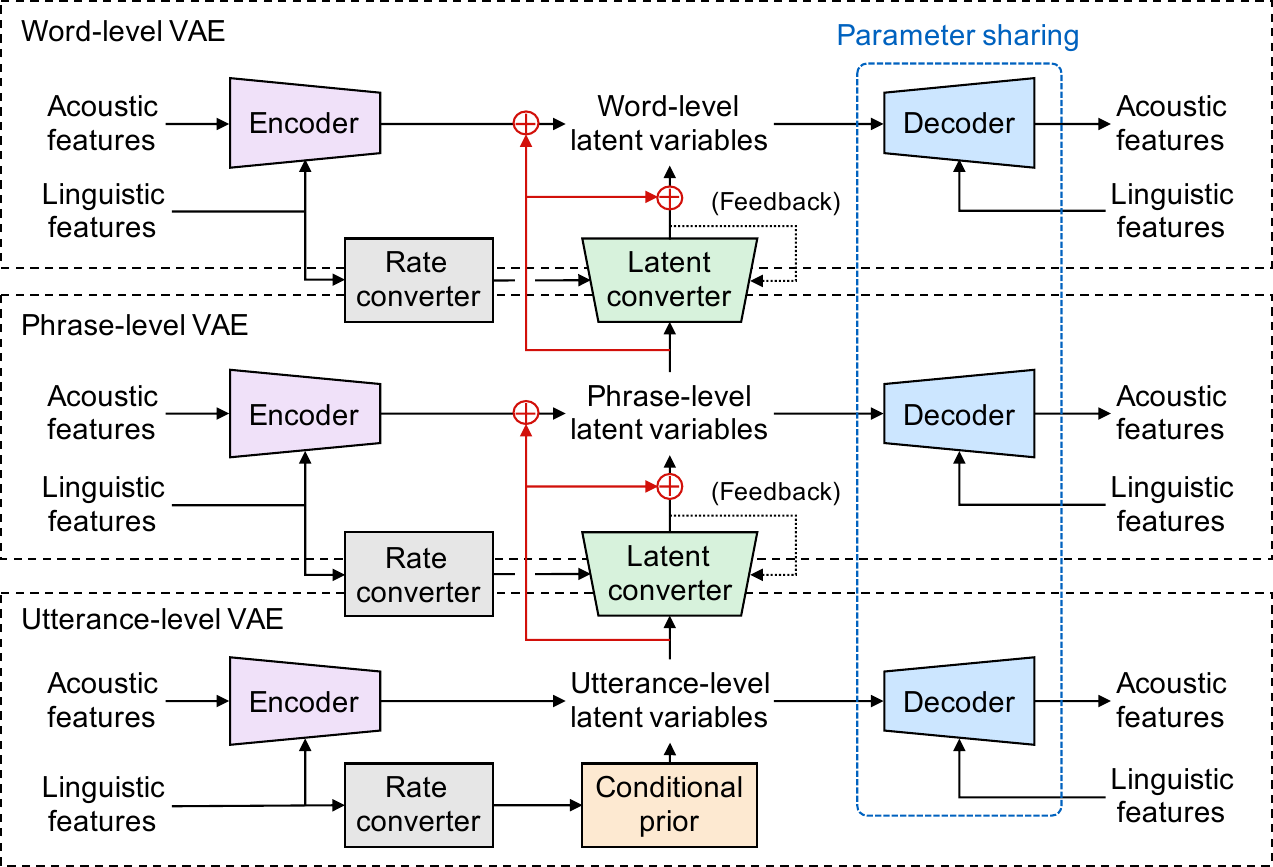}
  \vspace{-2mm}
  \caption{Overview of the proposed framework}
  \vspace{-4mm}
  \label{fig:proposed}
\end{figure}

An overview of the proposed framework is shown in Fig.~\ref{fig:proposed}.
This framework consists of a multi-grained VAE, a conditional prior, and a multi-level latent converter.
The multi-grained VAE is layered at three different time-resolution VAEs: utterance-level, phrase-level, and word-level, to extract each level latent representation.
The conditional prior is a distribution conditioned on the text, which enables sampling the latent variables by taking into account the content of the text.
The multi-level latent converter consists of two latent converters with AR structure: utterance-to-phrase and phrase-to-word latent converter.
Each latent converter predicts the finer-level latent variables from the coarser-level latent variables.
These converters can also be viewed as a conditional prior, which conditioned on not only the text but also the coarser-level latent variables.

In the proposed framework, the distribution of the conditional prior, utterance-to-phrase latent converter, and phrase-to-word latent converter are denoted as $p(\bz^u \mid \bY)$, $p(\bZ^p \mid \bz^u, \bY)$, and $p(\bZ^w \!\mid \bZ^p, \bY)$, respectively.
The joint probability of the latent variables can be written as
\begin{align}
  &p(\bZ^w, \bZ^p, \bz^u \mid \bY) = p(\bZ^w \!\mid \bZ^p, \bY) p(\bZ^p \mid \bz^u, \bY) p(\bz^u \mid \bY).
\end{align}
Therefore, the proposed framework can properly model the three dependencies of latent variables shown in Fig.~\ref{fig:z_graphical}.
Word-level latent variables are sampled from these distributions in a step-by-step manner, so fine-grained latent variables can be sampled by considering the hierarchical linguistic structure, the temporal coherency, and text content without any reference signals.
In addition, the encoders and the converters use residual connections to model the finer-level latent variables as variations from the coarser-level ones.
We also apply parameter sharing to all decoders so that all latent variables can share the same latent space.
We found that these are helpful for the latent converters to predict the finer-level latent variables from the coarser-level ones.

The encoders and the decoders have the same architecture described in Sec.~\ref{sec:vae-tts}.
The conditional prior has three FC layers including the output layer.
Each AR latent converter has one bidirectional LSTM layer and one unidirectional LSTM layer with an AR structure, which takes the previous output of the latent converter, and the FC layer as the output layer.
These conditional prior and latent converters take linguistic embeddings of each level.
To make these embeddings, we use rate converters with a stack of two bidirectional LSTM layers and a mean pooling layer.
We train this framework with two steps.
The first step involves training all encoders and decoders to learn each latent variable and predict the acoustic feature from the latent variable and linguistic feature.
The second step involves training the conditional prior and AR latent converters.
Scheduled sampling~\cite{bengio-2015-scheduled} is used for training the AR latent converters.

\section{Experiments}
\vspace{-1mm}
\subsection{Experimental conditions}
\vspace{-1mm}
In this experiment, a ten-hour Japanese single-female speaker corpus was used.
This corpus contains four speaking styles, i.e., normal (13,520 utterances), happy (3,861 utterances), sad (1,716 utterances), and radio (1,816 utterances).
Radio style is colloquial speech spoken in a radio program and includes different speaking styles within an utterance.
We used 20,073 utterances for training, 417 for validation, and 417 for testing.
The speech signals were sampled at 48 kHz, and each sample was quantized by 16 bits.
Feature vectors were extracted with a 5-ms shift, and the feature vector consisted of the $\log F_0$ value that is voting results from three $F_0$ estimators, the 0-th through 69-th WORLD mel-cepstral coefficients, and the 0-th through 34-th mel-cepstral analysis aperiodicity measures~\cite{Web-SPTK,morise-2016-world}.

Five-state, left-to-right, no-skip hidden semi-Markov models were used to obtain phoneme alignments~\cite{zen-2007-hidden}.
The phoneme durations were modeled using a style-dependent mixture density network~\cite{bishop-1994-mixture}.
The linguistic feature vector is a 718-dimensional vector, consisting of a full-context feature vector extracted by an external text analyzer~\cite{web-OpenJTalk} and duration feature vector including the duration of the current phoneme and the position of the current frame.
The acoustic feature vector is a 107-dimensional vector, consisting of 70-dimensional WORLD mel-cepstral coefficients, a $\log F_0$ value acquired by linearly interpolating in unvoiced parts, a voiced/unvoiced binary value, and 35-dimensional mel-cepstral analysis aperiodicity measures.
All latent variables in the VAEs are 2-dimensional to easily specify the value of latent variables.

\subsection{Reconstruction performance}
\vspace{-1mm}
\begin{table}[t]
  \caption{Objective evaluation of reconstruction performance}
  \label{tb:recon}
  \centering
  \vspace{-3mm}
  \begin{tabular}{lccc}
    \toprule
    Method & MCD & GVD & \hspace{-1mm}F0ER\hspace{-1mm} \\
    \midrule \midrule
    Utterance-level VAE (oracle $\bz^u$)\hspace{-1mm} & 5.023 & 0.541 & 0.113\\
    Word-level VAE (oracle $\bZ^w$) & 4.915 & 0.527 & 0.065\\
    \midrule
    \multicolumn{3}{l}{\hspace{-1.25ex}Multi-grained VAE  (predicted $\bZ^w$):} \\
    \textbf{M1} w/o residual \& dec. sharing\hspace{-1mm} & 5.048 & 0.538 & 0.117 \\
    \textbf{M2} w/ residual \& dec. sharing & 5.021 & 0.530 & 0.114 \\
    \bottomrule
  \end{tabular}
  \vspace{-4mm}
\end{table}

We calculated the objective scores to measure reconstruction performance.
Mel-cepstral distortion (MCD) [dB], global variance distance (GVD) for mel-cepstrum coefficients~\cite{hashimoto-2016-trajectory}, and root mean squared error of $\log F_0$ (F0ER) [logHz] were used.
The utterance-level VAE, word-level VAE, and two types of multi-grained VAE system were compared, i.e., \textbf{M1} denotes a multi-grained model without neither residual connection nor decoder's parameter sharing, and \textbf{M2} denotes a multi-grained model with residual structure and decoder parameter sharing.
In the utterance-level VAE and the word-level VAE, the latent variables are oracle ones extracted from natural acoustic features.
In \textbf{M1} and \textbf{M2}, the utterance-level latent variables are oracle ones, and the other level latent variables are predicted using the AR latent converters.

Table~\ref{tb:recon} shows the results of the objective evaluation.
Introducing the fine-grained latent variables improves the performance of predicting the acoustic features since these latent variables can capture the variations at a specific moment of speech.
Although \textbf{M1} was worse than the utterance-level VAE except regarding GVD, \textbf{M2} was improved and outperformed the utterance-level VAE in terms of MCD and GVD.
This result indicates that residual connection and decoder parameter sharing helps predict the finer-level latent variables.
However, \textbf{M2} did not reach the performance of word-level VAE, indicating that it is still challenging to predict the word-level latent variables.

\subsection{Subjective evaluation}
\vspace{-1mm}
To evaluate the naturalness and expressiveness of the synthesized speech, we conducted subjective listening tests.
The naturalness and expressiveness of the synthesized speech were assessed using mean opinion score (MOS) tests.
The opinion score in the MOS test for naturalness was based on a five-point scale (5: natural -- 1: poor in naturalness).
In the MOS test for expressiveness, a different five-point scale was used (5: very expressive -- 1: poor in expressiveness), and participants evaluate dit considering the content of the text.
The participants were 19 Japanese students in our research group, and 10 utterances were chosen at random per method from the test set.
For proper evaluation, we excluded short audio samples of less than 1 seconds except for the silence at both ends in advance.

The following six systems were compared.
\setlength{\leftmargini}{20pt}
\begin{itemize}
  \item \textbf{FG}:\;The fine-grained VAE-based system.
  The word-level latent variables were sampled from the normal Gaussian prior at the synthesis stage.
  \item \textbf{FG+AR}:\;The system in which the prior in \textbf{FG} is replaced by the AR prior.
  The AR prior consisted of one unidirectional LSTM layer and the FC layer as the output layer.
  \item \textbf{FG+CP}:\;The system in which the prior in \textbf{FG} is replaced by the conditional prior.
  This prior is a word-level conditional prior with the same architecture as the utterance-level conditional prior described in Sec.~\ref{sec:proposed}.
  \item \textbf{FG+CP+AR}:\;The system in which the prior in \textbf{FG} is replaced by the conditional AR prior.
  This prior was the network in which the feedback connection was added to the prior in \textbf{FG+CP}.
  \item \textbf{MG+CP}:\;The proposed system without the AR structure in the latent converter.
  At the synthesis stage, the utterance-level latent variable was sampled from the conditional prior, and the word-level latent variables were predicted by the non-AR latent converters.
  \item \textbf{MG+CP+AR}:\;The proposed system in which the latent converters in \textbf{MG+CP} are replaced with the AR latent converters.
\end{itemize}

\begin{table}[t]
  \caption{MOSs for naturalness and expressiveness}
  \label{tb:mos}
  \centering
  \vspace{-3mm}
  \begin{tabular}{lcc}
    \toprule
    Method & Naturalness & Expressiveness \\
    \midrule \midrule
    \textbf{FG}       & 3.20 $\pm$ 0.10 & 3.10 $\pm$ 0.10 \\
    \textbf{FG+AR}    & 3.20 $\pm$ 0.11 & 2.95 $\pm$ 0.10 \\
    \midrule
    \textbf{FG+CP}    & 3.29 $\pm$ 0.11 & 3.12 $\pm$ 0.12 \\
    \textbf{FG+CP+AR} & \textbf{3.47 $\pm$ 0.11} & 3.24 $\pm$ 0.11 \\
    \midrule
    \textbf{MG+CP}    & 3.44 $\pm$ 0.10 & 3.33 $\pm$ 0.11 \\
    \textbf{MG+CP+AR} & 3.31 $\pm$ 0.11 & \textbf{3.37 $\pm$ 0.11} \\
    \bottomrule
  \end{tabular}
  \vspace{-4mm}
\end{table}

Table~\ref{tb:mos} shows the result of subjective evaluation.
\textbf{FG+AR} had a worse score of expressiveness than \textbf{FG}.
Sampling from the AR prior is unstable because the latent variables after sampling are fed back to the prior network as the previous outputs.
\textbf{FG+CP} had a slightly better score than \textbf{FG}, but the difference between expressiveness was small.
We used the full-context feature vector as the linguistic feature vector, and the conditional prior took this vector.
This result suggests that it is difficult for the conditional prior to capture the meanings of the texts from the full-context features and sample the latent variables that match the content.
This may be improved by introducing a text-embedding model such as BERT~\cite{devlin-2018-bert,hayashi-2019-pre}.
Unlike between \textbf{FG} and \textbf{FG+CP}, \textbf{FG+CP+AR} outperformed \textbf{FG+CP}, which indicates that linguistic features help in the stable sampling of latent variables from the AR prior since full-context features include information about the linguistic structure.
Regarding the proposed systems, \textbf{MG+CP} and \textbf{MG+CP+AR} performed better than the others in terms of expressiveness.
This suggests that modeling the hierarchical linguistic structure of the latent variables is effective for expressive speech synthesis.
In addition, \textbf{MG+CP+AR} had the best score for expressiveness, so explicit temporal modeling of latent variables is useful.
Since the difference between them was small, modeling the hierarchical structure is more important than modeling the temporal coherency in latent variables.
On the other hand, the naturalness of \textbf{MG+CP} and \textbf{MG+CP+AR} was degraded compared to that of \textbf{FG+CP+AR}.
This may be caused by multiple sampling to obtain the word-level latent variables in the proposed system.

\subsection{Controllability}
\vspace{-1mm}
\begin{figure}[t]
  \centering
  \includegraphics[width=0.7\hsize]{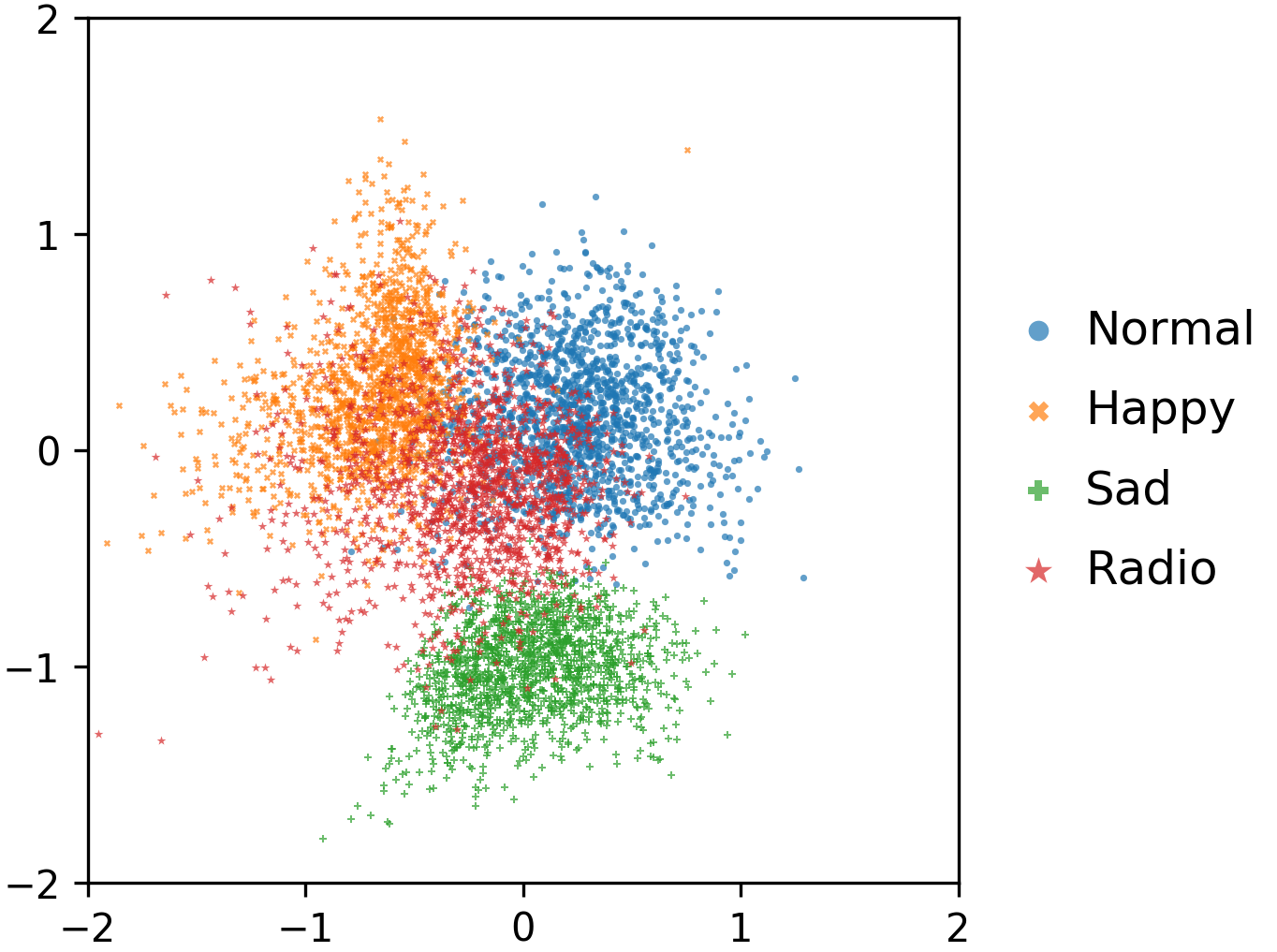}
  \vspace{-1mm}
  \caption{Latent space of the multi-grained VAE}
  \label{fig:z_utt_2d}
  \vspace{-6mm}
\end{figure}

Our proposed multi-grained system provides the controllability of speaking style at the synthesis stage.
The word-level latent variables are predicted from the utterance-level latent variables so that we can control the speaking style of the entire utterance despite using the fine-grained latent variables.
In this experiment, since the latent variable was set to 2 dimensions, we could visualize the utterance-level latent variables directly and manually specify their values at the synthesis stage.
Figure~\ref{fig:z_utt_2d} shows the latent space in the multi-grained VAE visualized by plotting the utterance-level latent variables of the randomly selected 1,500 utterances per speaking style.
The utterances of normal, happy, and sad style were classified without annotation, and the utterances of radio style, which includes various speaking styles, were distributed so as to be mixed into three styles.
Thus, we could control the speaking style of synthesized speech intuitively while viewing the latent space.
To show the controllability, we demonstrated synthesized speech samples on a web page\footnote{https://www.rinna.jp/research/interspeech2020/}.

\section{Conclusions}
\vspace{-1mm}
We proposed a novel framework for expressive speech synthesis with a hierarchical multi-grained generative model for modeling fine-grained latent variables, considering the hierarchical linguistic structure, the temporal coherency, and an input text.
Experimental results indicate that the proposed model is effective for expressive and controllable speech synthesis.
Future work includes utilizing a text-embedding model such as BERT to consider the content of the text in our proposed model.
Introducing the proposed model into end-to-end TTS with an attention mechanism is also included in our future work to control not only the acoustic features but also the phoneme durations.

\section{Acknowledgements}
\vspace{-1mm}
This work was supported by JSPS KAKENHI Grant Number JP19H04136.

\bibliographystyle{IEEEtran}
\bibliography{paper}

\begin{thebibliography}{10}
\providecommand{\url}[1]{#1}
\csname url@samestyle\endcsname
\providecommand{\newblock}{\relax}
\providecommand{\bibinfo}[2]{#2}
\providecommand{\BIBentrySTDinterwordspacing}{\spaceskip=0pt\relax}
\providecommand{\BIBentryALTinterwordstretchfactor}{4}
\providecommand{\BIBentryALTinterwordspacing}{\spaceskip=\fontdimen2\font plus
\BIBentryALTinterwordstretchfactor\fontdimen3\font minus
  \fontdimen4\font\relax}
\providecommand{\BIBforeignlanguage}[2]{{%
\expandafter\ifx\csname l@#1\endcsname\relax
\typeout{** WARNING: IEEEtran.bst: No hyphenation pattern has been}%
\typeout{** loaded for the language `#1'. Using the pattern for}%
\typeout{** the default language instead.}%
\else
\language=\csname l@#1\endcsname
\fi
#2}}
\providecommand{\BIBdecl}{\relax}
\BIBdecl

\bibitem{zen-2013-statistical}
H.~Zen, A.~Senior, and M.~Schuster, ``Statistical parametric speech synthesis
  using deep neural networks,'' in \emph{Proceedings of ICASSP}, 2013, pp.
  7962--7966.

\bibitem{fan-2014-tts}
Y.~Fan, Y.~Qian, F.-L. Xie, and F.~K. Soong, ``{TTS} synthesis with
  bidirectional {LSTM} based recurrent neural networks,'' in \emph{Proccdings
  of Interspeech}, 2014, pp. 964--1968.

\bibitem{wu-2016-inverstigating}
Z.~Wu and S.~King, ``Investigating gated recurrent networks for speech
  synthesis,'' in \emph{Proceedings of ICASSP}, 2016, pp. 5140--5144.

\bibitem{wang-2017-tacotron}
Y.~Wang, R.~Skerry-Ryan, D.~Stanton, Y.~Wu, R.~J. Weiss, N.~Jaitly, Z.~Yang,
  Y.~Xiao, Z.~Chen, S.~Bengio \emph{et~al.}, ``Tacotron: Towards end-to-end
  speech synthesis,'' in \emph{Proccdings of Interspeech}, 2017, pp.
  4004--4010.

\bibitem{shen-2018-natural}
J.~Shen, R.~Pang, R.~J. Weiss, M.~Schuster, N.~Jaitly, Z.~Yang, Z.~Chen,
  Y.~Zhang, Y.~Wang, R.~Skerrv-Ryan \emph{et~al.}, ``Natural {TTS} synthesis by
  conditioning wavenet on mel spectrogram predictions,'' in \emph{Proceedings
  of ICASSP}, 2018, pp. 4779--4783.

\bibitem{ping-2017-deep}
W.~Ping, K.~Peng, A.~Gibiansky, S.~O. Arik, A.~Kannan, S.~Narang, J.~Raiman,
  and J.~Miller, ``Deep voice 3: Scaling text-to-speech with convolutional
  sequence learning,'' \emph{arXiv preprint arXiv:1710.07654}, 2017.

\bibitem{li-2019-neural}
N.~Li, S.~Liu, Y.~Liu, S.~Zhao, and M.~Liu, ``Neural speech synthesis with
  transformer network,'' in \emph{Proceedings of the AAAI Conference on
  Artificial Intelligence}, vol.~33, 2019, pp. 6706--6713.

\bibitem{inoue-2017-investigation}
K.~Inoue, S.~Hara, M.~Abe, N.~Hojo, and Y.~Ijima, ``An investigation to
  transplant emotional expressions in {DNN}-based {TTS} synthesis,'' in
  \emph{Proceedings of APSIPA}, 2017, pp. 1253--1258.

\bibitem{skerry-2018-towards}
R.~Skerry-Ryan, E.~Battenberg, Y.~Xiao, Y.~Wang, D.~Stanton, J.~Shor, R.~J.
  Weiss, R.~Clark, and R.~A. Saurous, ``Towards end-to-end prosody transfer for
  expressive speech synthesis with tacotron,'' \emph{arXiv preprint
  arXiv:1803.09047}, 2018.

\bibitem{wang-2018-style}
Y.~Wang, D.~Stanton, Y.~Zhang, R.~Skerry-Ryan, E.~Battenberg, J.~Shor, Y.~Xiao,
  F.~Ren, Y.~Jia, and R.~A. Saurous, ``Style tokens: Unsupervised style
  modeling, control and transfer in end-to-end speech synthesis,'' \emph{arXiv
  preprint arXiv:1803.09017}, 2018.

\bibitem{kingma-2014-auto}
D.~P. Kingma and M.~Welling, ``Auto-encoding variational bayes.'' in
  \emph{Proceedings of ICLR}, 2014.

\bibitem{akuzawa-2018-expressive}
K.~Akuzawa, Y.~Iwasawa, and Y.~Matsuo, ``Expressive speech synthesis via
  modeling expressions with variational autoencoder,'' in \emph{Proccdings of
  Interspeech}, 2018, pp. 3067--3071.

\bibitem{henter-2018-deep}
G.~E. Henter, J.~Lorenzo-Trueba, X.~Wang, and J.~Yamagishi, ``Deep
  encoder-decoder models for unsupervised learning of controllable speech
  synthesis,'' \emph{arXiv preprint arXiv:1807.11470}, 2018.

\bibitem{zhang-2019-learning}
Y.-J. Zhang, S.~Pan, L.~He, and Z.-H. Ling, ``Learning latent representations
  for style control and transfer in end-to-end speech synthesis,'' in
  \emph{Proceedings of ICASSP}, 2019, pp. 6945--6949.

\bibitem{lee-2019-robust}
Y.~Lee and T.~Kim, ``Robust and fine-grained prosody control of end-to-end
  speech synthesis,'' in \emph{Proceedings of ICASSP}, 2019, pp. 5911--5915.

\bibitem{sun-2020-fully}
G.~{Sun}, Y.~{Zhang}, R.~J. {Weiss}, Y.~{Cao}, H.~{Zen}, and Y.~{Wu},
  ``Fully-hierarchical fine-grained prosody modeling for interpretable speech
  synthesis,'' in \emph{Proceedings of ICASSP}, 2020, pp. 6264--6268.

\bibitem{sun-2020-generating}
G.~{Sun}, Y.~{Zhang}, R.~J. {Weiss}, Y.~{Cao}, H.~{Zen}, A.~{Rosenberg},
  B.~{Ramabhadran}, and Y.~{Wu}, ``Generating diverse and natural
  text-to-speech samples using a quantized fine-grained vae and autoregressive
  prosody prior,'' in \emph{Proceedings of ICASSP}, 2020, pp. 6699--6703.

\bibitem{hochreiter-1997-long}
S.~Hochreiter and J.~Schmidhuber, ``Long short-term memory,'' \emph{Neural
  computation}, vol.~9, no.~8, pp. 1735--1780, 1997.

\bibitem{bengio-2015-scheduled}
S.~Bengio, O.~Vinyals, N.~Jaitly, and N.~Shazeer, ``Scheduled sampling for
  sequence prediction with recurrent neural networks,'' in \emph{Advances in
  Neural Information Processing Systems}, 2015, pp. 1171--1179.

\bibitem{Web-SPTK}
``Speech signal processing toolkit ({SPTK}),''
  \url{http://sp-tk.sourceforge.net/}.

\bibitem{morise-2016-world}
M.~Morise, F.~Yokomori, and K.~Ozawa, ``{WORLD}: a vocoder-based high-quality
  speech synthesis system for real-time applications,'' \emph{IEICE
  Transactions on Information and Systems}, vol.~99, no.~7, pp. 1877--1884,
  2016.

\bibitem{zen-2007-hidden}
H.~Zen, K.~Tokuda, T.~Masuko, T.~Kobayasih, and T.~Kitamura, ``A hidden
  semi-{M}arkov model-based speech synthesis system,'' \emph{IEICE Transactions
  on Information and Systems}, vol. E90-D, no.~5, pp. 825--834, 2007.

\bibitem{bishop-1994-mixture}
C.~M. Bishop, ``Mixture density networks,'' Neural Computing Research Group,
  Aston University, Tech. Rep. NCRG/94/004, 1994.

\bibitem{web-OpenJTalk}
``{O}pen {JT}alk,'' \url{http://open-jtalk.sourceforge.net/}.

\bibitem{hashimoto-2016-trajectory}
K.~Hashimoto, K.~Oura, Y.~Nankaku, and K.~Tokuda, ``Trajectory training
  considering global variance for speech synthesis based on neural networks,''
  in \emph{Proceedings of ICASSP}, 2016, pp. 5600--5604.

\bibitem{devlin-2018-bert}
J.~Devlin, M.-W. Chang, K.~Lee, and K.~Toutanova, ``{BERT}: Pre-training of
  deep bidirectional transformers for language understanding,'' in
  \emph{Proceedings of NAACL-HLT}, 2019.

\bibitem{hayashi-2019-pre}
T.~Hayashi, S.~Watanabe, T.~Toda, K.~Takeda, S.~Toshniwal, and K.~Livescu,
  ``Pre-trained text embeddings for enhanced text-to-speech synthesis,'' in
  \emph{Proccdings of Interspeech}, 2019, pp. 4430--4434.

\end{thebibliography}

\end{document}